\documentclass[conference]{IEEEtran}
\ifCLASSINFOpdf
\else
\fi
\usepackage{soul}

\usepackage[cmex10]{amsmath}

\usepackage{graphicx}
\usepackage{epstopdf}
\usepackage{amsfonts}
\usepackage{xcolor}
\makeatletter
\def\testclr#1#{\@testclr{#1}}
\def\@testclr#1#2{{\fboxsep\z@\fbox{\colorbox#1{#2}{\phantom{XX}}}}}

\makeatother

\definecolor{babyblue}{rgb}{0.54, 0.81, 0.94}
\begin{document}
\title{Sliding Window Spectrum Sensing for Full-Duplex Cognitive Radios with Low Access-Latency}


\author{Orion~Afisiadis, Andrew~C.~M.~Austin, Alexios~Balatsoukas-Stimming
        and Andreas~Burg% <-this % stops a space
\thanks{The authors are with the Telecommunications Circuits Laboratory, 
\'{E}cole Polytechnique F\'{e}d\'{e}rale de Lausanne, Lausanne, Switzerland.}}% <-this % stops a space

\linespread{1.00}

\maketitle

\begin{abstract}

In a cognitive radio system the failure of secondary user (SU) transceivers to promptly vacate the channel 
can introduce significant \emph{access-latency} for primary or high-priority users (PU). In conventional cognitive radio systems, the backoff latency is exacerbated by frame structures that only allow sensing at periodic intervals. Concurrent transmission and sensing using self-interference suppression has been suggested to improve the performance of cognitive radio systems, allowing decisions to be taken at multiple points within the frame. In this paper, we extend this approach by proposing a sliding-window full-duplex model allowing decisions to be taken on a sample-by-sample basis. We also derive the access-latency for both the existing and the proposed schemes. Our results show that the access-latency of the sliding scheme is decreased by a factor of $2.6$ compared to the existing slotted full-duplex scheme and by a factor of approximately $16$ compared to a half-duplex cognitive radio system. Moreover, the proposed scheme is significantly more resilient to the destructive effects of residual self-interference compared to previous approaches.

\end{abstract}


\IEEEpeerreviewmaketitle

\section{Introduction}

One of the most significant challenges faced by wireless systems today is the (apparent) scarcity of available spectrum. However, recent studies have shown that while the radio spectrum is densely allocated, it is often not heavily occupied or utilized by the licensed primary users (PU)~\cite{FCCreport2002,Chiang2007}. Frequency-agile cognitive radio networks have been proposed to take advantage of this situation, by allowing unlicensed secondary users (SU) to opportunistically reuse licensed frequency bands~\cite{MitolaIII1999}. One of the fundamental requirements of such systems is that SUs should not generate harmful interference to PUs. Consequently, SU transceivers must be capable of sensing the radio channel to determine if a PU is present~\cite{Haykin2005}. 
Similar to spectrum re-use, we can also consider a scenario that requires low-latency medium access for high-priority 
users or high-priority transmissions. In such systems, a less latency-sensitive ongoing transmission would need to stop instantly as a PU forcefully accesses the channel to get its urgent message across. Such scenarios are of particular importance for real-time services, such as virtual/augmented reality and autonomous vehicles and therefore the proposed standards for future 5G systems aim to achieve less than 1~ms latency~\cite{GSM5G2014}. 

System architectures that periodically stop SU transmissions to sense the channel have been widely proposed~\cite{Yucek2009,Liang2008} to detect the start of PU transmissions. These approaches introduce \emph{blind-intervals}, where the SU system is transmitting and thus unable to detect the start of a PU transmission until the next sensing slot (at the earliest). Decreasing the interval between successive sensing-slots will decrease the efficiency and throughput of the SU system~\cite{Liang2008} but improve its spectrum sensing capabilities. Previous research has examined the trade-off between sensing and throughput, and introduced scheduling algorithms to maximise sensing efficiency~\cite{Lee2008}, and MAC-layer frame structures to maximise SU throughput while adequately protecting the PU~\cite{Liang2008}. 
An alternative approach to improve the detection-throughput trade-off, by using self-interference cancellation to enable concurrent transmission and sensing (similar to full-duplex systems~\cite{Sabharwal2014}), has been proposed by a number of authors~\cite{Ahmed2012,Afifi2013,Riihonen2014,Afifi2014,YunLiao2014}. Unfortunately, in practice, perfect self-interference suppression is usually not attained, and for typical operating conditions the residual self-interference component is above the noise floor~\cite{Balatsoukas2015}. Previous research has considered the analysis of the sensing-throughput trade-off for energy detection~\cite{Afifi2013}, waveform detection~\cite{Afifi2015}, the power-throughput trade-off~\cite{YunLiao2014,YunLiao2015} and various adaptive algorithms for maximizing SU throughput~\cite{Riihonen2014,Afifi2014}, in the presence of different levels of residual self-interference. 

%It should be noted that in practise, perfect self-interference cancellation (i.e., suppression below the thermal noise floor) is usually not attained, due to transmit non-linearities, phase noise and uncertainties in the channel estimates

In existing work, the PU is assumed to be protected if the probability of detection, $P_d$, is sufficiently high (typically above 90\%)~\cite{Yucek2009,Afifi2014}. However, from a physical-layer perspective, the \emph{access-latency} (defined as the time required by the SU to detect the PU and vacate the channel) gives a better measure of the impact of SU interference and the necessary protection. For example, a high access-latency may harm PU communication, by distorting a too large portion of its training or synchronisation fields. Moreover, the full-duplex\footnote{
In this paper the conventional cognitive radio system is referred to as a half-duplex scheme, while all the systems using self-interference suppression to enable concurrent transmission and sensing are referred to as full-duplex schemes.
} cognitive radio schemes considered in~\cite{Afifi2013} and~\cite{YunLiao2014} typically take decisions after $N_s$ samples have been accumulated into a buffer. These schemes can thus have a high access-latency, as the PU may start transmitting at any time.  

\emph{Contributions:} In this paper we focus on obtaining analytical expressions for the physical access-latency in various cognitive radio systems. The access-latency results are presented in terms of physical samples and can thus be scaled and applied to arbitrary hardware implementations. Furthermore, to alleviate the issue of high access-latency in full-duplex cognitive radio systems, we introduce a full-duplex \emph{sliding-window} spectrum sensing technique. Unlike existing schemes, our approach takes decisions on a sample-by-sample basis, and can detect the presence of PUs more quickly, thereby reducing the access-latency as demonstrated by our simulation results. 

\emph{Outline:} This paper is organized as follows. In Section II we describe the system models for the existing half-duplex and slotted full-duplex systems, along with the model of our proposed sliding-window full-duplex system. The throughput-latency trade-off is analysed in Section III, along with derivations of expressions for the access-latency. Theoretical and numerical simulation results are shown in Section IV, and we conclude in Section V.

%\begin{itemize}
%	\item We introduce a \emph{sliding window} approach for spectrum sensing in cognitive radio with self-interference suppression capabilities. We call this method \emph{sliding full-duplex}.
%	\item We derive \emph{analytical formulas} of the access-latency for the conventional half-duplex, the existing slotted full-duplex and the sliding full-duplex schemes.
%	\item The results show that the sliding model can achieve \emph{lower latency} compared to both half-duplex and slotted full-duplex systems.
%	\item Residual self-interference deteriorates the performance of both full-duplex systems. The \emph{relative} deterioration for the sliding system is smaller though, making it more resistant to residual self-interference.
%	\item We show that using full-duplex cognitive radio is \emph{meaningful} (from a latency-throughput perspective) for both models, compared to the half-duplex case.
%	
%\end{itemize}
%Examining the back-off latency issue for cognitive radio. Protection of the PU in existing bibliography is measured with a high Pd. We examine the protection of PU with the actual number of samples until the SU backs-off. More meaningful approach for the physical layer.

\section{System Models}

	\begin{figure}
		\centering
		\includegraphics[width=0.47\textwidth]{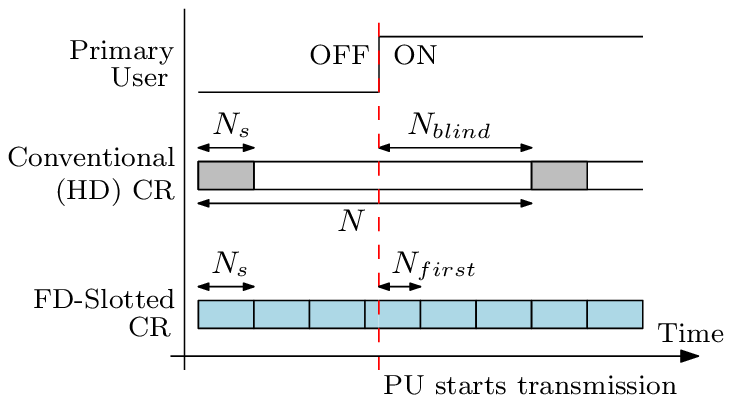} 
		\caption[Frame Structures]{Frame structures for conventional (HD) and concurrent (FD) cognitive radio systems:~\testclr{white}~: SU transmission;
		\testclr{gray!50!white}~: sensing only; and \testclr{babyblue}~: concurrent transmit and sense. }
%		 \includegraphics[width=0.47\textwidth]{Figs/block_diagram2.eps}
% 		\caption[Frame Structures]{Frame structures for conventional (HD) and concurrent (FD) cognitive radio systems:~\testclr{white}~: SU transmission;
% 		\testclr{gray!50!white}~: sensing only; \testclr{babyblue}~: concurrent transmit and sense; \testclr{violet}~: sliding-window concurrent transmit and sense. }
		\label{fig:blockdiagram}
	\end{figure}

In this section we outline the three cognitive radio system models considered in this paper. 
For all three approaches, the presence of the PU is detected by comparing a decision metric, $M$, 
(computed from a set of $N_s$ samples), against a threshold, $\epsilon$. If $M>\epsilon$, 
the PU is assumed to be present, otherwise the channel is declared idle.   
A widely used decision-metric for spectrum sensing is energy detection, with~\cite{Yucek2009}
\begin{equation}
	M = \frac{1}{N_{s}} \sum_{n=1}^{N_{s}} |r(n)|^{2},
	\label{eq.energy}
\end{equation}
where $r(n)$ is the received signal. Other metrics to detect the presence of the PU are not considered in this paper, however, the findings can be readily extended.
%Another more sophisticated method, which takes known patterns in the PU signal into account, is waveform 
%detection, with a decision metric given by~\cite{}
%\begin{equation}
%	M = \Re \left[\sum_{n=1}^{N_{s}} r(n)x^{*}(n)\right],
%	\label{eq.waveform}
%\end{equation}
%where $x(n)$ is part of the known PU signal, e.g., a preamble. Other approaches to
%detect the presence of the PU are not considered in this paper.

Based on the decision metric, the probabilities of detection, $P_d$, and false-alarm, $P_f$, can be found by applying appropriate 
hypothesis tests~\cite{Yucek2009,Liang2008}. Denoting $ \mathcal{H}_{0} $ as the scenario where the PU is inactive, 
and $ \mathcal{H}_{1} $ where the PU is active, the probabilities of false alarm and detection are defined as
\begin{align}
	 P_{f} &= Pr(M > \epsilon | \mathcal{H}_{0}) \label{eq.pf} \\
	 P_{d} &= Pr(M > \epsilon | \mathcal{H}_{1}). \label{eq.pd}
\end{align} 
The probability of misdetection, $ P_{m} $ is defined as
\begin{equation}
	 P_{m} = Pr(M < \epsilon|\mathcal{H}_{1}) = 1 - P_{d}.	\label{eq.pm}
\end{equation}	 

\subsection{Half-Duplex Cognitive Radio}

Fig.~\ref{fig:blockdiagram} shows the SU transmission frame consisting of $N$ samples for a conventional half-duplex cognitive 
radio system~\cite{Liang2008}. The frame consists of two parts: an $N_s$ sample sensing window and an 
$N - N_s$ sample transmission window. The $N_s$ samples are used to decide if the PU is present based on the decision metric (e.g.,~(\ref{eq.energy})), with transmission continuing if the PU is not detected. Decisions are thus made every $N$ samples (i.e., once per frame). It is important to note that the SU is unable to detect or react to 
the presence of the PU while transmitting, leading to a high access-latency. Due to the sensing window, the throughput of the SU system is also reduced by a factor of $\frac{N-N_s}{N}$.  

\subsection{Concurrent Sensing and Transmission With Slotted Sensing}

By cancelling the self-interference signal, SUs can concurrently transmit and sense the channel. Fig.~\ref{fig:blockdiagram} shows the frame structure of a concurrent sensing and transmission system using self-interference 
cancellation as proposed in~\cite{Afifi2015} and~\cite{YunLiao2014}. The systems analysed in~\cite{Afifi2015} retain the same frame structure as a 
conventional CR system by including a sensing-only slot at the start of each frame. In this paper, we consider full-duplex systems 
that only include concurrent transmission and sensing slots, similar to the approach taken in~\cite{YunLiao2014,YunLiao2015}.  
The SU throughput is not reduced as there is no dedicated sensing-slot. Similarly, the SU may be able to more 
quickly detect whether the PU starts transmitting during the frame, thereby reducing the access-latency. While decisions are made more 
frequently (every $N_s$ samples) than in a conventional cognitive radio system (every $N$ samples), there still remains a blind interval of $N_{\text{first}}$ samples. Residual self-interference also makes it more difficult to detect the presence of the PU, by effectively increasing the noise-floor. However, as the detection decisions are made more frequently compared to the half-duplex case, lower values of the $P_d$ may still result in an acceptable access latency.   

\subsection{Concurrent Sensing and Transmission with Sliding Window Sensing}

We propose an extension to the concurrent sensing and transmission scheme outlined in the previous section by introducing a sliding-window to take decisions at every sample, i.e., the buffer does not wait to fill with $ N_{s} $ fresh samples before a decision is made. 
This approach can be implemented easily in digital hardware via a FIFO buffer. It is important to note that successive 
decision metrics are \emph{not} independent as only one new sample is added (and one removed). However, the minimum access-latency of this scheme is one sample, unlike the slotted full-duplex method.  In our analysis it is assumed that the number of sensing samples, $ N_{s} $, remains the same as the slotted full-duplex model. 

%% AA: this needs to be rewritten completely, very unclear what this is trying to say
%In this subsection, we introduce a new sliding model for the full-duplex SU. In this sliding model the number of sensing samples, $ N_{s} $, stays the same as the slotted full-duplex model. The main difference of the new sliding model is that a decision is made every time a new sample is available, i.e. in every clock. Thus, the buffer does not wait to fill with $ N_{s} $ fresh samples before a decision is made. The first sample of the buffer is discarded and it is replaced, in a fifo way, by the one that just arrived. This way, in each new decision, only 2 samples are actually different from the previous decision. Thus the sensing metric in each new decision is \emph{not} independent from the metric that was used in the previous decision. The total times a decision is taken is $ N_{s} $ times larger than the slotted full-duplex model and $ N $ times larger than the half-duplex model. Thus, a potentially lower $ P_{d} $ for the same performance, or equivalently a better performance for the same $ P_{d} $, comes again as an argument to test.

\section{Latency-Throughput analysis}

\subsection{Throughput of the Half-Duplex and Full-Duplex Systems}

The capacity of the SU (assuming the PU is not transmitting) is given by
\begin{equation}
	C_{0} = \log_{2}\left(1 + \text{SNR}_{\text{SU}}\right),
\end{equation} where $\text{SNR}_{\text{SU}} $ is the signal-to-noise ratio of the SU measured at a receiver node.
This capacity is achieved when the PU is not transmitting \emph{and} the SU has not raised a false alarm. 

Following~\cite{Liang2008,YunLiao2014}, we express the total throughput for the SU system with full-duplex sensing 
capability as
\begin{equation}
	R_{\text{FD}} = C_{0}(1 - P_{f}).
\end{equation} 
For a specific SNR of the SU, the throughput is a parameter of only $P_f$. In practice, if a false-alarm occurs, the loss in SU throughput can be significant as an entire data frame may be lost, however the impact of these intermittent SU outages can be mitigated by coding. 

The half-duplex system has an additional throughput loss due to the sensing overhead
\begin{equation}
	R_{\text{HD}} = \frac{N - N_{s}}{N}C_{0}(1 - P_{f}).
	\label{eq:HD_througputreduction}
\end{equation}

Clearly for the half-duplex system there exists a trade-off between detection latency and throughput that is determined by the sensing overhead. For full-duplex systems, the main consideration is the increase of $ P_{f} $, due to the residual self-interference. In order for a full-duplex system with residual self-interference to maintain a throughput close to $ C_{0} $, it has to operate with lower $ P_{d} $ values. In the results we show that the sliding window model allows for lower access-latency compared to the slotted full-duplex and half-duplex cases, while maintaining a high SU throughput.

%From the SU's perspective only $ P_{f} $ matters; the lower it is the better for the SU throughput.
%Since in existing bibliography, protection of the primary user is measured with the value of $ P_{d} $, %the higher the $ P_{d} $ is the  better PU is considered to be protected. But as $ P_{d} $ gets higher, %so does $ P_{f} $, a fact that  leads to throughput loss.
%We consider protection of the PU by the actual number of samples needed for the SU to back-off from the %moment the PU turns on. We call this the \emph{back-off latency}. This approach is more appropriate from %a physical layer point of view. A PU will give the constraint by the actual time (samples) that he can %afford to be interfered by the SU. 
%Decision is taken every sample, thus we expect that in the FD models, and especially the sliding one, %these extra opportunities for the SU to detect the PU will provide a lower latency for the same %throughput or a higher throughput for the same latency. Thus a better latency-throughput trade-off.

\subsection{Average Latency of the Half-Duplex System}

%In this subsection we will derive the average access-latency in samples for the half-duplex system. Of course, in order to derive the access-latency, we suppose that the PU is off and that the SU is transmitting.

Let us denote $ D_{k} $ as the event that the PU is detected during the $ k $-th decision after starting transmission, and $ D_{k}^{\mathsf{c}} $ the complementary event that the PU is not detected during the $ k $-th decision. Let also $ N_{k} $ be the number of samples from when the PU starts to the $ k $-th decision point. The average access-latency for all the schemes can thus be computed from the infinite sum
\begin{align}
	L =& N_{1}P(D_{1}) + N_{2}P(D_{2}|D_{1}^{\mathsf{c}}) + N_{3}P(D_{3}|D_{1}^{\mathsf{c}} \cap D_{2}^{\mathsf{c}}) + \nonumber 
	\\ &\ldots + N_{i}P(D_{i}|D_{1}^{\mathsf{c}}\cap D_{2}^{\mathsf{c}} \ldots \cap D_{i-1}^{\mathsf{c}}) + \ldots  
	\label{eq:GeneralLatency}
\end{align}

For the half-duplex case, there are two possible scenarios, depending on which part of the SU activity (sensing or transmitting) the PU starts transmitting. In the first scenario the PU starts transmitting during the blind interval, $ N_\text{blind} $ samples before the end of the blind interval. The first decision is made after $ (N_\text{blind} + N_{s}) $ samples, i.e., $ N_{1} = (N_\text{blind} + N_{s}) $. Let $ P_{d}(k) $ denote the probability of detection with $k$ samples of the PU signal and $ P_{m}(k) $ denote the probability of misdetection for the same case; the probability of detecting the PU at the first decision is thus $ P_{d}(N_{s}) $. If the PU is not detected, the number of samples for the second decision will be $ N_{2} = (N_\text{blind} + N_{s} + N) $, again with probability of detection $ P_{d}(N_{s}) $, and so on. Thus
\begin{align}
	L(N_\text{blind}) =& P_{d}(N_{s})\sum_{n=0}^{\infty}(N_\text{blind} + N_{s} + nN)P_{m}^{n}(N_{s}) \nonumber 
	% \\ =& P_{d}(N_{s})\left[(N_{blind} + N_{s})\sum_{n=0}^{\infty}P_{m}^{n}(N_{s}) + \right. \nonumber 
	% \\ & \left. N\sum_{n=0}^{\infty}nP_{m}^{n}(N_{s})\right] \nonumber 
	\\ =& P_{d}(N_{s})\left[(N_\text{blind} + N_{s})\frac{1}{P_{d}(N_{s})} + N\frac{P_{m}(N_{s})}{P_{d}^{2}(N_{s})}\right],
\end{align} for $ 0 < P_{d}(N_{s}) < 1 $. 

$ N_\text{blind} $ can take each value between $ 1 $ and $ N - N_{s} $ with probability $ \dfrac{1}{N} $, thus the average latency under scenario $ 1 $ is
\begin{equation}
	L_{1} = \dfrac{1}{N}\sum_{N_\text{blind}=1}^{N-N_{s}}{L(N_\text{blind})}.
\end{equation}

In the second scenario the PU starts transmitting during the sensing period, $ N_\text{first} $ samples before the end of the sensing period. The first decision is taken after $ N_{1} = N_\text{first} $ samples, with probability of detection $ P_{d}(N_\text{first}) $. If the PU is not detected, the number of samples for the second decision is $ N_{2} = (N_\text{first} + N) $, with probability of detection $ P_{d}(N_{s}) $, and so on. Thus 
\begin{align}
	L(N_\text{first}) =& N_\text{first}P_{d}(N_\text{first}) + P_{d}(N_{s})P_{m}(N_\text{first})\times \nonumber
	\\ & \sum_{n=0}^{\infty}\left[N_\text{first} + (n+1)N\right]P_{m}^{n}(N_{s}) \nonumber
%	\\ =& N_{first}P_{d}(N_{first}) + P_{d}(N_{s})P_{m}(N_{first})\left[N_{first}\sum_{n=0}^{\infty}P_{m}^{n}(N_{s}) + N\sum_{n=0}^{\infty}(n+1)P_{m}^{n}(N_{s})\right] \nonumber
	\\ =& N_\text{first}P_{d}(N_\text{first}) + P_{d}(N_{s})P_{m}(N_\text{first})\times \nonumber
	\\ & \left[N_\text{first}\frac{1}{P_{d}(N_{s})} + N\frac{1}{P_{d}^{2}(N_{s})}\right]
	\label{eq:hd_latency}
\end{align} for $ 0 < P_{d}(N_{s}) < 1 $. 

As $ N_\text{first} $ can take any value between $ 1 $ and $ N_{s} $, with probability $ \dfrac{1}{N} $, the average latency under scenario~2 is
\begin{equation}
	L_{2} = \dfrac{1}{N}\sum_{N_\text{first}=1}^{N_{s}}{L(N_\text{first})}.
\end{equation} 
Therefore, the total average latency of the half-duplex system is $ L = L_{1} + L_{2} $.

\subsection{Average Latency of the Slotted Full-Duplex System} 

In the slotted full-duplex scheme there is no blind interval. Let us thus consider the case where the PU turns on $N_{\text{first}}$ samples before the end of the sensing period. The first decision is made after $ N_{1} = N_\text{first} $ samples, with probability of detection $ P_{d}(N_\text{first}) $. If the PU is not detected, the second decision has latency of $ N_{2} = (N_\text{first} + N_{s}) $, with $ P_{d}(N_{s}) $, and so on. Thus the latency is 	
\begin{align}
	L(N_\text{first}) =&  N_\text{first}P_{d}(N_\text{first}) + P_{d}(N_{s})P_{m}(N_\text{first}) \times \nonumber
	\\ &\sum_{n=0}^{\infty}\left[N_\text{first} + (n+1)N_{s}\right]P_{m}^{n}(N_{s}) \nonumber
%	\\ =& N_{first}P_{d}(N_{first}) + P_{d}(N_{s})P_{m}(N_{first}) \times \nonumber
%	\\ & \left[ N_{first}\sum_{n=0}^{\infty}P_{m}^{n}(N_{s}) + N_{s}\sum_{n=0}^{\infty}(n+1)P_{m}^{n}(N_{s}) \right] \nonumber
	\\  =&  N_\text{first}P_{d}(N_\text{first}) + P_{d}(N_{s})P_{m}(N_\text{first}) \times \nonumber
	\\ & \left[N_\text{first}\frac{1}{P_{d}(N_{s})} + N_{s}\frac{1}{P_{d}^{2}(N_{s})} \right],
	\label{eq:fd_slotted_latency}
\end{align} 
for $ 0 < P_{d}(N_{s}) < 1 $. 

$ N_\text{first} $ can take any value between $ 1 $ and $ N_{s} $ with probability $ \dfrac{1}{N_{s}} $, which leads to an average latency
\begin{equation}
	L = \dfrac{1}{N_{s}}\sum_{N_\text{first}=1}^{N_{s}}{L(N_\text{first})}.
\end{equation}

\subsection{Average Latency of the Sliding Full-Duplex System} \label{sec:latency_sliding}

In the sliding full-duplex model, the average latency can be derived from (\ref{eq:GeneralLatency}), however the difference in this case is that the decisions are not independent. A decision taken at a sample $ i $ is not independent of decisions taken over the previous $ i + (N_{s}-1) $ samples. Only decisions separated by $ N_{s} $ samples are independent. Accordingly, the slotted model can be regarded as a special case of the sliding model, where the only decisions kept are those separated by $ N_{s} $ samples. The idea of the sliding approach is that there is no real reason to discard all the decisions in between, as these will potentially reduce the access-latency. It should also be noted that taking decisions every sample is feasible in contemporary systems. However, the effect on the energy consumption, which may be a limitation for battery-driven SUs, remains to be investigated.
%It should also be noted that taking decisions every sample is feasible in contemporary systems and does not add additional complexity to the SU transceiver, though it may increase energy consumption, which may be a limitation for power-limited battery-driven SUs. The performance metrics of the scheme with respect to computational load and energy consumption are currently being investigated.

Since each decision is not independent of the $ N_{s} $ previous decisions, the conditional terms of (\ref{eq:GeneralLatency}) cannot be 
easily expanded using multiplicative terms as in~(\ref{eq:hd_latency}) and~(\ref{eq:fd_slotted_latency}). The analytical expression for the average access-latency in this case would require the conditional terms to be expressed using a stochastic process model with memory, and is outside the scope of this work. To compare the performance of our proposed sliding-window method with the existing schemes, we have used Monte-Carlo simulations which maintain the dependence between the decisions.

\section{Results and Discussion}

\begin{figure}
		\centering
		\includegraphics[width=0.5\textwidth]{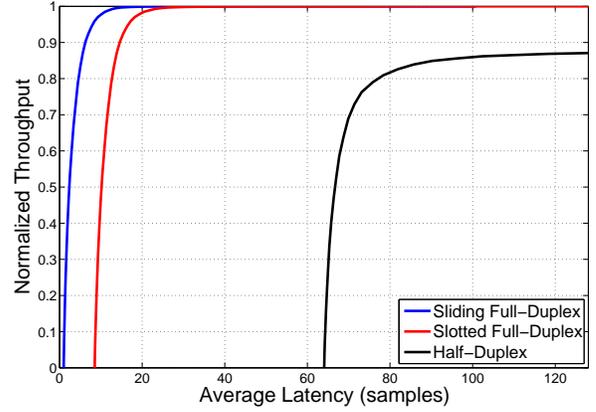} 
		\caption{Average latency - normalized throughput comparison for the 3 systems, $\text{SNR}_\text{PU} = 0$ dB with perfect self-interference suppression, $ N_{s} = 16 $, $ N = 128 $.}
		\label{fig:AvgLatency}
		\vspace{-2mm}
	\end{figure}

In Fig.~\ref{fig:AvgLatency} we observe a significant improvement in the average latency of the full-duplex schemes compared to the conventional half-duplex scheme. Moreover an improvement is observed for the sliding full-duplex model compared to the slotted scheme. For these results, no residual self-interference is present, i.e., perfect self-interference suppression is assumed. The minimum average access-latency of the half-duplex scheme is approximately half the length of a SU frame, as the system is unable to sense while transmitting. The slotted full-duplex system has lower access-latency---as decisions are made every $N_s$ samples---and the minimum average latency is approximately $\frac{N_s}{2}$. The sliding-window full-duplex scheme can potentially detect the PU with an average access-latency of a single sample, however, this mode of operation results in very low SU throughput (and is thus not viable in normal circumstances). Similarly, the maximum throughput is achieved for all three schemes---the half-duplex system has a reduced maximum throughput due to the sensing overhead, as given by~(\ref{eq:HD_througputreduction})---as the average latency increases. 

% AA: ??
%These improvements can be interpreted in two equivalent ways: 
%\begin{itemize}
%\item  A lower latency for the same throughput: Same throughput means the same $ P_{f} $ value for all systems, thus also the same $ P_{d} $ value. The system that takes more decisions will have a lower latency.
%\item A higher throughput for the same latency: Same latency for the sliding model is interpreted in a lower accepted $ P_{d} $ value, a fact that leads also to a lower $ P_{f} $ value, thus a higher throughput.
%\end{itemize} 

		\begin{figure}
		\centering
		\includegraphics[width=0.5\textwidth]{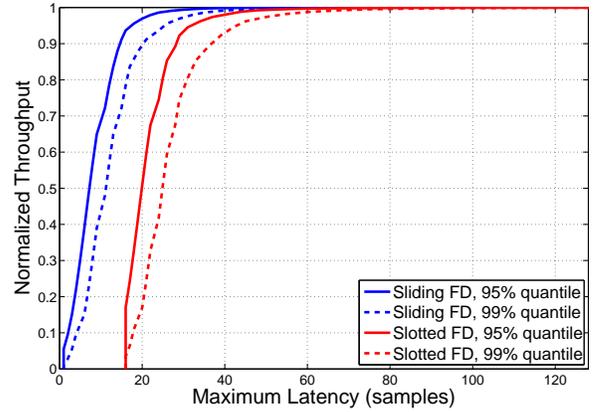} 
		\caption{Maximum acceptable latency - normalized throughput comparison for the $95$\% and $99$\% quantiles for the 3 systems, primary user $\text{SNR}_\text{PU} = 0$, $ N_{s} = 16 $, $ N = 128 $.}
		\label{fig:Quantiles}
		\vspace{-2mm}
	\end{figure}

{Fig.~\ref{fig:Quantiles} shows the $95$\% and $99$\% quantiles of the latency-throughput curves for both full-duplex schemes. The access-latency is a random variable, which depends on the actual realisations of the signal, noise and self-interference. Thus, in order to provide useful metrics of the system performance we examine the maximum access-latency that is reached with a specific confidence.

%% AA: I have no idea what any of this means: see paragraph above for my interpretation
%Fig.~\ref{fig:Quantiles} shows the guaranteed maximum latency the PU can afford for a specific percentage of the time.
%We depict two strict quantiles at 90\% and 95\% for the sliding and the slotted full-duplex models. Thus a significant improvement is observed also on the maximum acceptable access-latency for the sliding model, compared to the slotted one. The conventional half-duplex model is not depicted since its latency is too large compared to both the full-duplex systems.
%Give an example like: For a maximum latency of 40 samples for 90 percent of the time, the slotted scheme can give a normalized throughput of 0.5, while the sliding scheme gives a throughput over 0.95. There is a significant throughput gain for the SU for the same protection of the PU.
\begin{figure}
	\centering
 	\includegraphics[width=0.5\textwidth]{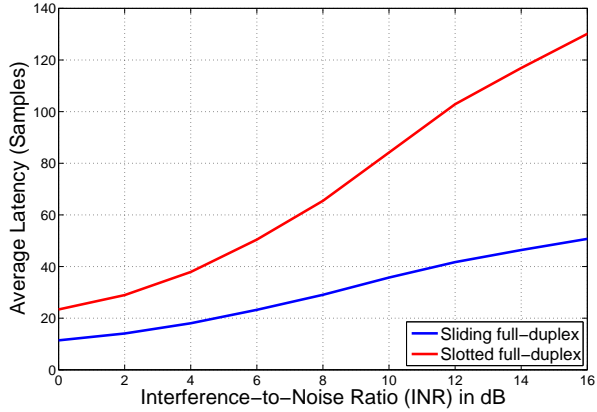}
 	\caption{Average access-latency with increasing residual self-interference,  for a normalised throughput of 0.9, $\text{SNR}_\text{PU} = 0$~dB, $ N_{s} = 16 $, $ N = 128 $.}
 	\label{fig:INR}
 	\vspace{-2mm}
\end{figure}
\begin{figure}
	\centering
	\includegraphics[width=0.5\textwidth]{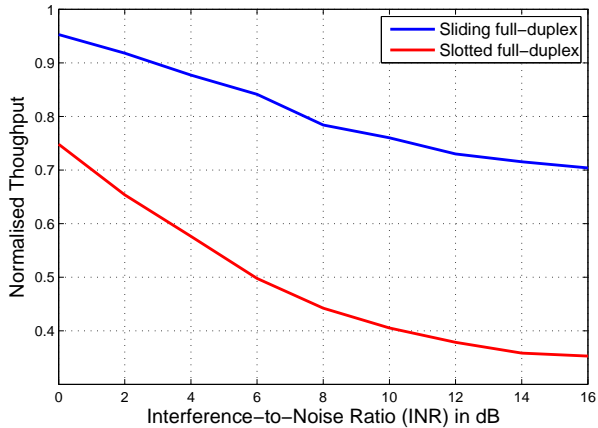}
	\caption{Normalised throughput with increasing residual self-interference, for an average access-latency of 16 samples, $\text{SNR}_\text{PU} = 0$~dB, $ N_{s} = 16 $, $ N = 128 $.}
	\label{fig:INR2}
	\vspace{-2mm}
\end{figure}

Fig.~\ref{fig:INR} compares the average latency for the two full-duplex models with increasing residual self-interference (measured relatively to the noise-floor). The average latency is found by normalizing the throughput to $0.9$,  i.e., $ P_{f} = 0.1 $. The sliding scheme is observed to have lower latency for all values of residual self-interference considered. Moreover, the difference in the slopes indicates that the sliding scheme is more resistant to residual self-interference, by approximately a factor of $2.7$ relative to the slotted scheme.

In Fig.~\ref{fig:INR2} we can observe a similar result from a different perspective. In this figure we compare the normalised throughput for the two full-duplex models with increasing residual self-interference. The throughput is found for the same latency of $ L=N_{s} $ samples for both schemes, i.e., for the same PU protection. The sliding scheme is observed to have higher throughput for all values of residual self-interference considered, again with a difference in the slopes.

\section{Conclusions}
In this paper we have elaborated on the analysis of cognitive radios from the perspective of protecting the primary or high-priority users by reducing the access-latency of the system. By deriving analytical formulas for the access-latency of existing schemes, both half-duplex and full-duplex, we have quantified the access-latency problem, and showed that it is particularly significant for the half-duplex model and that it is reduced by the slotted full-duplex model. In order to overcome the latency problem even more effectively, we proposed a sliding full-duplex scheme that profits from its ability to take decisions every sample.  The results show that there is a significant improvement in the response performance of the system by using full-duplex techniques. The problem of residual self-interference that exists in every full-duplex system was also considered. The proposed sliding method was proven to resist to the residual self-interference 2.7 times more effectively than the slotted scheme.

%It has been also shown that the proposed scheme is about 2 times more resistant to residual self-interference compared to the existing full-duplex model.


\bibliographystyle{IEEEtran}
\bibliography{refs}

\end{document}